\documentclass{aa}
\usepackage{amsmath}
\graphicspath{{./}{./fig/}}

\usepackage{graphicx}
\usepackage{txfonts}

\begin{document}

\title{Formation of giant planets with large metal masses and metal fractions via giant impacts in a rapidly dissipating disk}
\titlerunning{Origins of HD\,149026b and TOI-849b}

\author{Masahiro Ogihara\inst{1}
        \and
        Yasunori Hori\inst{1,2}
        \and
        Masanobu Kunitomo\inst{3}
        \and
        Kenji Kurosaki\inst{4}
          }
\authorrunning{Ogihara et al.}
   
\institute{National Astronomical Observatory of Japan, 2-21-1, Osawa, Mitaka, 181-8588 Tokyo, Japan
\email{ogihara@nagoya-u.jp}
\and
Astrobiology Center, 2-21-1, Osawa, Mitaka, 181-8588 Tokyo, Japan
\and
Kurume University, 67 Asahimachi, Kurume, 830-0011 Fukuoka, Japan
\and
Nagoya University, Furo-cho, Chikusa-ku, Nagoya, 464-8602 Aichi, Japan
}
   
\abstract
{According to planetary interior models, some giant planets contain large metal masses with large metal-mass fractions. HD\,149026b and TOI-849b are characteristic examples of these giant planets.
It has been suggested that the envelope mass loss during giant impacts plays a key role in the formation of such giant planets.
}
{The aim of the present letter is to propose a mechanism that can explain the origin of such giant planets.}
{We investigate the formation of giant planets in a rapidly dissipating disk using \textit{N}-body simulations that consider pebble accretion.}
{The results show that although the pebble isolation mass is smaller than the metal mass ($\gtrsim 30$ Earth masses) in some giant planets, the interior metal mass can be increased by giant impacts between planets with the isolation mass. 
Regarding the metal fraction, the cores accrete massive envelopes by runaway gas accretion during the disk-dissipation phase of 1--10\,Myr in a disk that evolves without photoevaporation.
Although a large fraction of the envelope can be lost during giant impacts,
the planets can reaccrete the envelope after impacts in a slowly dissipating disk.
Here, we demonstrate that, by photoevaporation in a rapidly dissipating disk, the runaway gas accretion is quenched in the middle, resulting in the formation of giant planets with large metal-mass fractions.}
{The origins of HD\,149026b and TOI-849b, which are characterized by their large metal-mass fractions, can be naturally explained by a model that considers a disk evolving with photoevaporation.}

\keywords{Planets and satellites: formation -- Planets and satellites: atmospheres -- Planets and satellites: individual: HD 149026b -- Planets and satellites: individual: TOI-849b}
\maketitle               

\section{Introduction} \label{sec:intro}

HD\,149026b is considered a hot Jupiter with a mass of $M = 117 \,M_\oplus$ \citep{2005ApJ...633..465S}. This planet likely has a large metal (or heavy element) mass of $>40 \,M_\oplus$ \citep{2009ApJ...695L.159D,2009ApJ...696..241C}. Recently, the TESS mission discovered a planet, TOI-849b, with $M = 39 \,M_\oplus$, which is mostly composed of heavy elements \citep{2020Natur.583...39A}.
The heavy element mass in giant planets has also been estimated for other planets, whose masses and radii are both measured.
\citet{2016ApJ...831...64T} derived the metal mass of giant planets (mostly warm Jupiters) and found that the typical metal mass range is approximately 10--100 $M_\oplus$; they also found that the metal-mass fraction is typically higher than 0.1.

Although giant planets with large heavy-element masses ($>30 \,M_\oplus$) appear to be relatively common in observations, the accretion of large heavy-element masses is difficult when planetary cores are being assembled. When cores grow by accreting surrounding planetesimals, their growth ceases at the planetesimal isolation mass ($= 2 \pi a \Delta a \Sigma_{\rm d}$). Assuming $\Delta a =10 R_{\rm H}$, where $R_{\rm H}$ is the Hill radius, an extremely high solid surface density of $\Sigma_{\rm d}\simeq 1000 {\rm \,g \,cm^{-2}}$ is required for the formation of cores larger than $30 \,M_\oplus$ at a semimajor axis of $a = $0.3 {\rm \,au}. Even if pebble accretion is considered, the pebble isolation mass is usually smaller than $25 \,M_\oplus$ \citep[e.g.,][]{2014A&A...572A..35L}.
It is possible that giant planets that formed in wide orbits ($a >$ a few tens of au) accrete planetesimals of several tens of Earth masses during migration \citep{2020A&A...633A..33S}; however, the formation of giant planets in such wide orbits is challenging \citep{2019A&A...623A..88B,2019A&A...631A..70J}.

Additionally, even if massive planetary cores are available in protoplanetary disks, subsequent runaway gas accretion typically does not yield giant planets with metal fractions higher than 10\%. Giant planets are formed by the accretion of the gaseous components of the protoplanetary disk. Since the accretion process proceeds in a runaway fashion \citep[e.g.,][]{1996Icar..124...62P}, giant planets accrete massive envelopes, in which case the metal-mass fraction is low.
For the formation of giant planets with high metal-mass fractions, the photoevaporation of accreted envelopes may play a role.
The origin of HD\,149026b and TOI-849b, however, is unlikely to be explained by the photoevaporation process because massive cores of $\gtrsim 30\,M_\oplus$ should suppress the envelope loss by the energy-limited hydrodynamic escape \citep[e.g.,][]{2014ApJ...783...54K}. The interior modeling suggests that HD\,149026b has an envelope of $\gtrsim$ 30\% of its total mass \citep{2006ApJ...642..495F,2006ApJ...650.1150I}, whereas the envelope fraction of TOI-849b is no more than 3.9wt\%. If TOI-849b had a larger radius (about 1.5 times the Jupiter radius), the Roche-lobe overflow driven by tidal disruption would be a possible mechanism for the origin of the large metal fraction \citep[e.g.,][]{2020Natur.583...39A}.
As an alternative way, the envelope mass loss during giant impacts can play an important role. For example, \citet{2006ApJ...650.1150I} claim that a substantial number ($\sim 90\%$) of envelopes can be lost by a high-speed ($\sim$2.5 times the escape velocity) collision between giant planets, which is considered a possible mechanism for the formation of HD\,149026b. For the origin of TOI-849b, \citet{2020Natur.583...39A} also suggest that the envelope can be removed during the giant planet collision, enabling the formation of planets with large metal-mass fractions.

Meanwhile, \citet{2020ApJ...899...91O} investigated the formation of super-Earths and sub-Neptunes (SENs) and demonstrated that SENs with low-mass atmospheres were formed in a rapidly dissipating disk by photoevaporation. We expect that such a disk would also help explain the origin of giant planets with high metal fractions.

\section{Model description}\label{sec:model}

We investigate the planet formation using a similar approach to that in \citet{2020ApJ...899...91O}, to which we refer the readers for full details. Several salient points are briefly described here.

The key ingredient of our model is the detailed disk evolution. \citet{2016A&A...596A..74S} developed a 1D disk-evolution model that considers mass loss due to magnetohydrodynamic (MHD) disk winds \citep{Suzuki+Inutsuka09} and wind-driven accretion \citep{1982MNRAS.199..883B}, in addition to viscous evolution. Subsequently, \citet{2020MNRAS.492.3849K} improved this model by considering the mass loss due to photoevaporative winds (PEWs) \citep[e.g.,][]{Alexander+06a}. The temporal evolution of the gas-surface density is shown in Figure~1 of \citet{2020ApJ...899...91O}. We observe that the surface density profile is flattened in the close-in region inside $r \sim 1 {\rm \,au}$, where $r$ is the radial distance from the star due to the effects of the disk winds. The bottom panels of Figure~\ref{fig:t_a} in this letter show the disk accretion rate of $r = 0.3 {\rm \,au}$ for models with and without PEWs. The disk-accretion rate is as high as $100 M_\oplus/{\rm Myr}$ even at $t = 10 {\rm \,Myr}$ for models with no PEWs, while for models with PEWs, the disk-accretion rate rapidly decreases during the disk-dissipation phase ($t > 1 {\rm \,Myr}$).

Another major physical ingredient is the envelope mass loss during giant impacts. \citet{2014LPI....45.2869S} derived a scaling law for the mass loss during giant impacts \citep[see Section~2.1.6 of][]{2020ApJ...892..124O}; we utilized an updated model of \citet{2020ApJ...901L..31K}. Although the mass loss was estimated by the ground speed in \citet{2014LPI....45.2869S}, the envelope was directly included in the simulations of \citet{2020ApJ...901L..31K}. The envelope loss fraction, $X_1$, from the planet with $M_{\rm core,1}$ in a collision with a planet with $M_{\rm core,2}$ is given by
\begin{eqnarray}
X_1 = 0.64 \left[\left(\frac{v_{\rm imp}}{v_{\rm esc}}\right)^{2} \left(\frac{M_{\rm core,1}}{M_{\rm tot}}\right)^{1/2} 
f_M\right]^{0.65},
\end{eqnarray}
where $v_{\rm imp}$, $v_{\rm esc}$, and $M_{\rm tot}$ are the impact velocity, two-body escape velocity, and total core mass ($=M_{\rm core,1} + M_{\rm core,2}$), respectively. The fractional interacting mass, $f_M$, is given by Eq.~(B1) of \citet{2020ApJ...901L..31K}. Here, we assume that the bulk core densities of colliding planets are equal. The envelope loss from the planet with $M_{\rm core,2}$ was calculated using the same method.

Here, we summarize other conditions. The simulations were initiated with 40 bare embryos with $M = 0.01 \,M_\oplus$, which were placed between $a = 0.2$--$1 {\rm \,au}$ with orbital separations of 15 mutual Hill radii. Cores can grow via pebble accretion \citep[e.g.,][]{2010A&A...520A..43O,2012A&A...544A..32L} until they reach the pebble isolation mass \citep[e.g.,][]{2018A&A...612A..30B}. As mentioned in Section~\ref{sec:intro}, the pebble isolation mass is smaller than $25 \,M_\oplus$; thus, it is difficult to explain the massive metal mass of the observed giant planets. The pebble flux is assumed to be $3 \times 10^{-4} \exp(-t/{\rm Myr})\,M_\oplus\,{\rm yr}^{-1}$, which is larger than that used in the simulations of super-Earth formation \citep{2020ApJ...899...91O}, to form more massive planets. The pebble flux was reduced in accordance with the number of pebbles accreted on outer planets.

Cores can start accreting massive envelopes when they reach the critical core mass \citep{2020ApJ...892..124O}:
\begin{eqnarray}\label{eq:miso}
M_{\rm crit} = 13 \,M_\oplus \left( \frac{\dot{M}_{\rm acc,pb}}{10^{-6} \,M_\oplus \,{\rm yr^{-1}}}\right)^{0.23},
\end{eqnarray}
where $\dot{M}_{\rm acc,pb}$ denotes the pebble accretion rate. While pebble accretion results in core growth, it can also heat the envelope, leading to a significant increase in the critical core mass. When the pebble accretion ceases after the cores reach the pebble isolation mass, the critical core mass decreases to a value lower than $10\,M_\oplus$ \citep{2010ApJ...714.1343H}. 
The actual envelope accretion rate was calculated by the Kelvin-Helmholtz contraction and regulated by the disk-accretion rate, $\dot{M}_{\rm disk}$.

Orbits of planets with masses larger than that of Mars change due to the gravitational interaction with the disk gas. We consider the damping of eccentricity and inclination \citep[e.g.,][]{2008A&A...482..677C}, as well as orbital migration \citep[e.g.,][]{2011MNRAS.410..293P,2018ApJ...861..140K}.

\section{Results}\label{sec:results}

\begin{figure*}
\centering
\includegraphics[width=1.6\columnwidth]{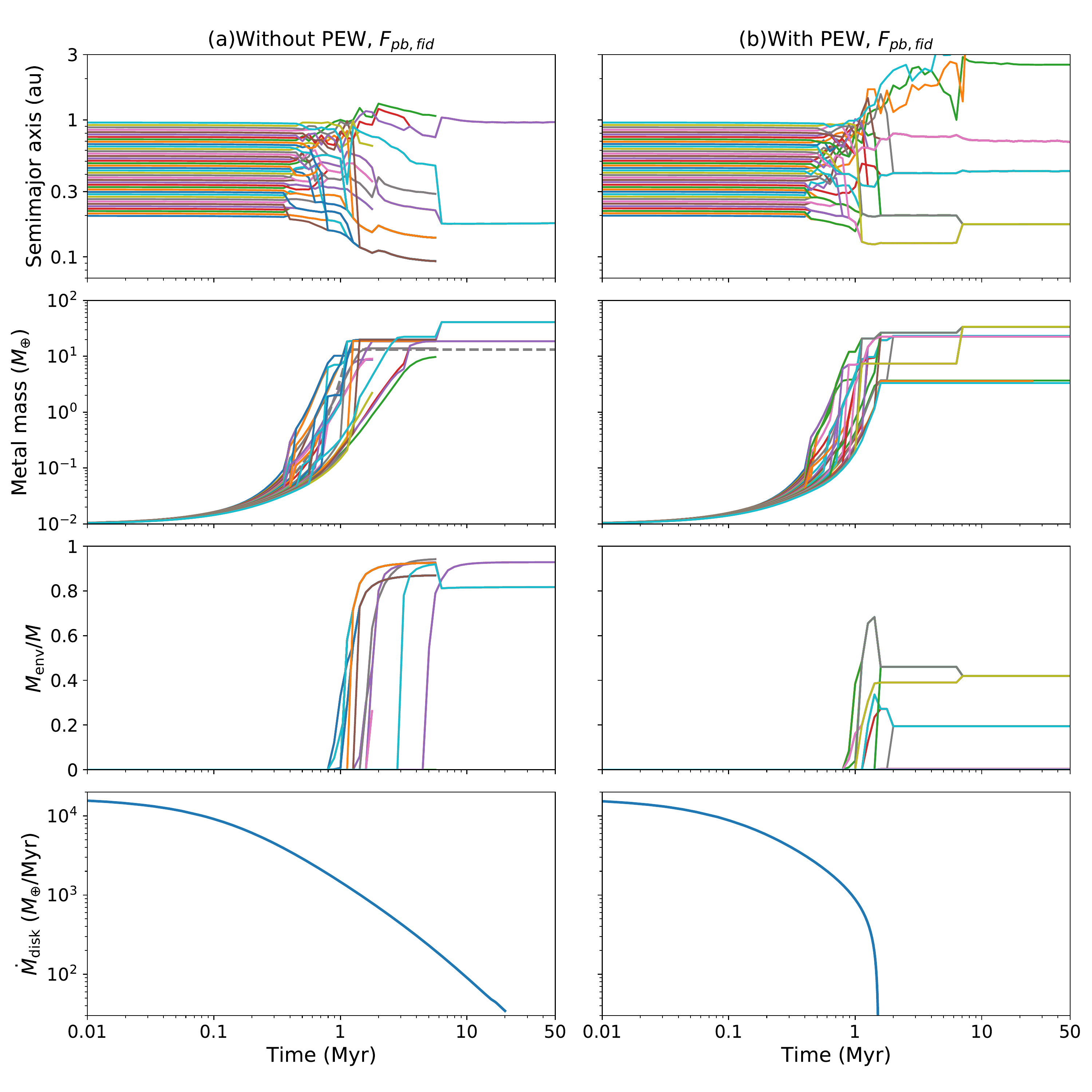}
\caption{Evolution of the semimajor axis $a$, metal mass $M_{\rm z}$, and envelope mass fraction $M_{\rm env}/M$ for our fiducial model in disks with and without PEW effects. The dashed line in panel (a) indicates that the growth of the core ceases at the pebble isolation mass if mutual collisions are ignored. The bottom panels show the gas accretion rate at $r = 0.3 {\rm \,au}$.}
\label{fig:t_a}
\end{figure*}

We present the typical simulation results. Figure~\ref{fig:t_a} shows the temporal evolution of the semimajor axis, the metal mass, and the envelope mass fraction for simulations with our fiducial pebble flux. Panel~(a) indicates the results of simulations for a disk without PEWs, while panel~(b) corresponds to those of simulations for a disk with PEWs. As mentioned in Section~\ref{sec:model}, the bottom panels show the disk-accretion rate at $r = 0.3\,{\rm au}$.
First, we examined the results of simulations in a disk with no PEWs (panel (a)). As shown in the evolution of the semimajor axis, planets did not undergo significant migration\footnote{According to the suppression of type-I migration \citep[e.g.,][]{2018A&A...615A..63O, 2020ApJ...899...91O}, it is possible to prevent super-Earth cores from having an icy composition, which is consistent with the inference that the majority of super-Earths and sub-Neptunes have rocky cores \citep[e.g.,][]{2021PSJ.....2....1A, 2020arXiv200711006R}.}. This is because the gas-surface density profile was altered from simple power-law distributions due to the effects of mass loss and mass accretion caused by MHD disk winds.
Further, we find that planets exhibit orbital instability (hence the giant impact events) during the disk dissipation ($t \gtrsim 1\,{\rm Myr}$). After giant impact events and the ejection of some planets, including the two innermost planets, a few giant planets eventually formed.

The second row of Figure~\ref{fig:t_a} highlights the planets that can have a large mass of heavy elements ($> 10 \,M_\oplus$). The gray dashed curve indicates the metal mass evolution of planets at $a = 0.3$\,au for a reference simulation in which the mutual interactions and collisions between planets were ignored. This shows that the growth of cores is terminated when they reach the pebble isolation mass ($\simeq 10$--$20 \,M_\oplus$). Conversely, in the results of \textit{N}-body simulations that consider mutual interactions, planets undergo giant impacts after they reach the pebble isolation mass. Therefore, here, we demonstrate that the high metal mass of giant planets can be achieved by giant impacts. Similar results promoting the formation of massive cores have also been demonstrated in \textit{N}-body simulations of cold-Jupiter formation \citep{2020MNRAS.496.3314W}. A related analytical argument has also been reported by \citet{2020MNRAS.498..680G} recently.

Regarding the envelope mass fraction, planets acquire massive envelopes ($>80 {\rm wt\%}$) in a disk with no PEW effects. While planets grow via pebble accretion, the critical core mass can increase to a value larger than $10\,M_\oplus$ due to the heating of the envelope by the accreting pebbles (Eq.\,\ref{eq:miso}). After planets reach the pebble isolation mass, the critical core mass decreases to $\simeq 3\,M_\oplus$ or smaller \citep{2010ApJ...714.1343H}. At $t \simeq 1 {\rm \,Myr}$, the cores exceed the critical core mass and start accreting massive envelopes. The gas-accretion rate in the disk is sufficiently high for envelope accretion ($>100 \,M_\oplus \,{\rm Myr}^{-1}$) until $t \simeq 10 {\rm \,Myr}$ (see bottom panel), leading to the accretion of massive envelopes. Moreover, we find that even if a portion of the envelope is lost during giant impacts, the envelope mass fraction is high at the end. This is because planets can reaccrete envelopes after envelope erosion during impacts, which is discussed in Section~\ref{sec:erosion}.

Here, we examine the results of simulations in a disk with PEW effects (panel (b)). The evolutions of the semimajor axis and metal mass are similar to those in panel (a). In other words, planets do not undergo significant migration and have a high mass of heavy elements after giant impacts. In the evolution of the envelope mass fraction, we observe a notable difference. Planets can avoid accreting massive envelopes, and the final envelope mass fraction is less than 0.5. This is because the gas disk is rapidly cleared out during the disk-dissipation phase ($t\gtrsim 1 {\rm \,Myr}$), and the envelope accretion onto planets is quenched. Consequently, we find that both the high metal mass and the high metal fraction in the observed giant planets can be explained in a rapidly dissipating disk with PEWs\footnote{There is a slight difference in the number of remaining planets between Figure~\ref{fig:t_a}(a) and (b). This difference may be due to the presence of runaway gas accretion. In all of our simulations, however, there is no clear trend that the number of remaining planets depends on whether or not PEWs are included.}.

\begin{figure*}[ht!]
\centering
\includegraphics[width=1.6\columnwidth]{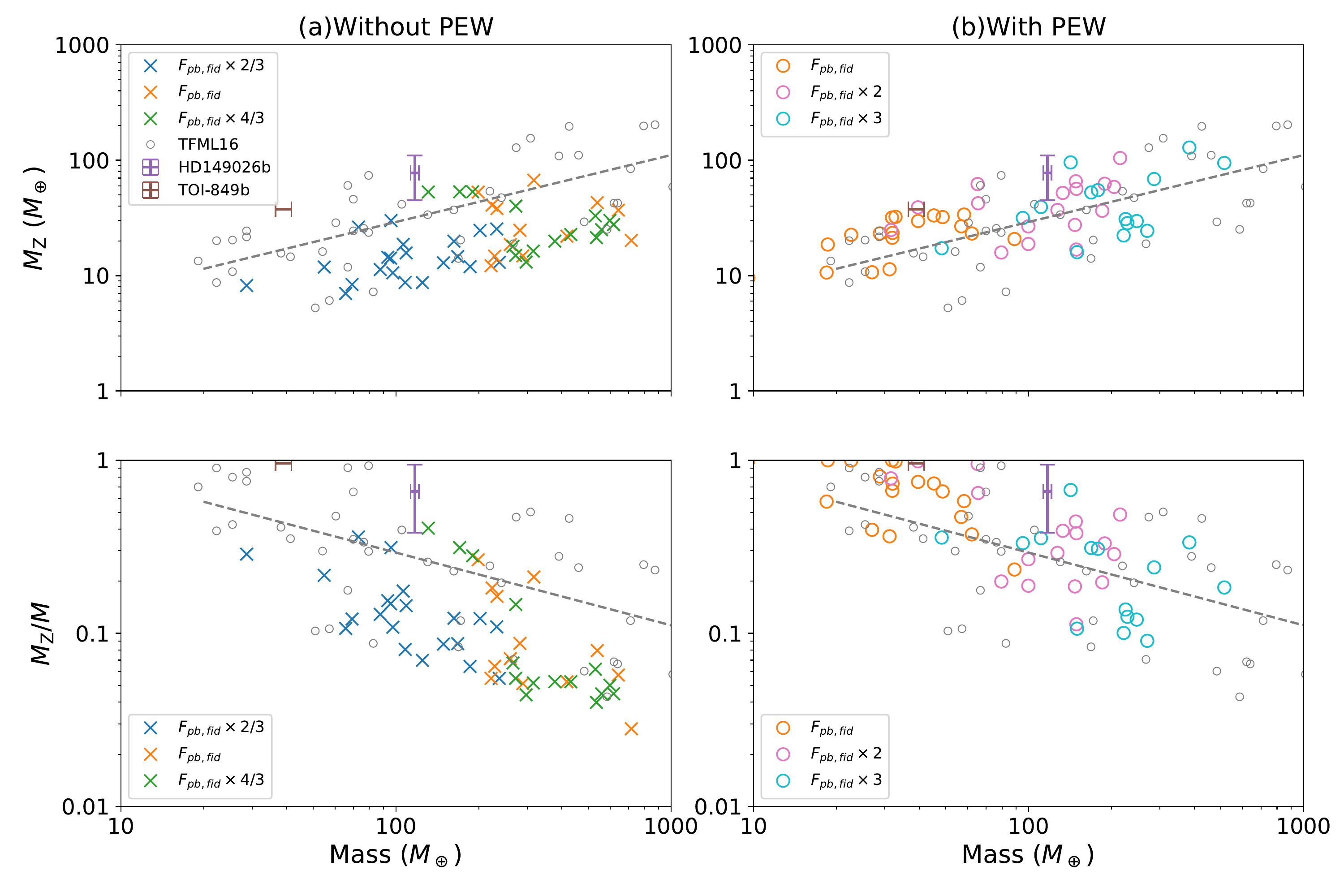}
\caption{Comparison of the metal mass $M_{\rm z}$ and metal-mass fraction $M_{\rm z}/M$ between the observed transiting giant planets and the results of our simulations. Gray circles indicate the estimates by \citet{2016ApJ...831...64T}, which were fitted by power-law distributions (dashed lines). HD\,149026b and TOI-849b are plotted with purple and brown symbols, respectively.
The simulation results are plotted with cresses and circles. The pebble flux was varied, and five runs were performed for each pebble flux.}
\label{fig:m_mcore}
\end{figure*}

Figure~\ref{fig:m_mcore} compares the estimates of the metal mass of the observed giant planets with our simulation results. The gray circles represent the estimates derived by \citet{2016ApJ...831...64T}. Both the metal mass and the metal-mass fraction can be fitted by simple power laws. HD\,149026b and TOI-849b are also plotted with purple and brown symbols, respectively. Panels (a) and (b) indicate the results of simulations in disks without and with PEW effects, respectively. The pebble flux was varied from the fiducial flux of $F_{\rm pb,fid}=3 \times 10^{-4} \exp(-t/{\rm Myr})\,M_\oplus\,{\rm yr}^{-1}$, in which the results of simulations with the same pebble flux are indicated with the same color. The values of the pebble flux were chosen such that the final masses of the planets are between $\simeq$ 20 and 1000${\rm \,M_\oplus}$. For the cases without PEWs, the pebble flux was varied between 2/3 and $4/3 \,F_{\rm pb,fid}$ (the time-integrated mass flux is between 200 and 400$\,M_\oplus$). For the cases with PEWs, the flux was varied between 1 and $3 \,F_{\rm pb,fid}$ (the time-integrated mass flux is between about 230 and 690$\,M_\oplus$). Although the total mass flux is high, some observations suggest a high solid mass reservoir in protoplanetary disks \citep[e.g.,][]{2016ApJ...821L..16C}.
Five runs with different initial locations of embryos were performed for each pebble flux.
Regardless of whether the PEW effects are considered or not, it is shown that planets with a large metal mass of more than $30 \,M_\oplus$ can form. This is due to the promoted growth by giant impacts between the planets, which stop their growth once the pebble isolation mass is achieved.

Regarding the metal-mass fraction, the fraction tends to be lower than 0.1 after the accretion of massive envelops in a disk with no PEW effects, which is generally lower than the estimates of \citet{2016ApJ...831...64T}. 
For the simulations in a disk with PEW effects, on the other hand, the metal-mass fraction tends to be higher than 0.1 because the envelope accretion is terminated by the rapid disk clearing. The metal-mass fraction is consistent with the estimates for observed giant planets. 
The trend that massive planets tend to have lower metal-mass fractions is interpreted as follows. For relatively large pebble fluxes, cores reach the pebble isolation mass quite early and start the runaway gas accretion. In addition, large cores can form after giant impacts of large pebble fluxes.

Furthermore, we find that the origins of the distinctive planets (i.e., HD\,149026b and TOI-849b) can be explained by our simulations. In the simulations with no PEW effects, no planets that are similar to the distinctive planets form. However, we find that, in a disk with rapid disk clearing by PEWs, planets with a high metal mass and metal fraction similar to HD\,149026b and TOI-849b can be formed. Although the origins of the large metal fraction of these two planets are not well understood (see Section~\ref{sec:intro}), our model can explain their origins. We note that the formation of hot Jupiters ($a < 0.1 {\rm\,au}$) was not directly treated in our simulations because of the extreme computational cost required for \text{N}-body simulations of close-in planets. According to previous simulations that investigate the formation of closer-in planets \citep[][]{2020ApJ...892..124O}, it is most likely that orbital locations of hot Jupiters can also be explained by our model.

More interestingly, the relationship between the eccentricity and the metal mass suggested by \citet{2016ApJ...831...64T} also seems to be better explained with PEWs. Figure 13 of \citet{2016ApJ...831...64T} suggests that planets with a high eccentricity tend to have a high heavy-element mass, although this is not statistically significant because the number of planets is not high. We do not go into detail here; however, we find that this trend can be better explained in the case with PEWs.

\section{Effects of envelope erosion during giant impacts}\label{sec:erosion}

For the origin of HD\,149026b and TOI-849b, it has been suggested that the impact erosion of accreted envelopes during a     high-speed impact $(>2.5 \,v_{\rm esc})$ between planets plays an important role in increasing the metal-mass fraction. 
In the present study, however, we find that the impact erosion of envelopes alone does not explain the formation of giant planets with low envelope mass fractions. We instead show that the cutoff in disk accretion by PEWs plays a crucial role. We discuss this point thoroughly.

\begin{figure}[ht!]
\centering
\includegraphics[width=0.9\columnwidth]{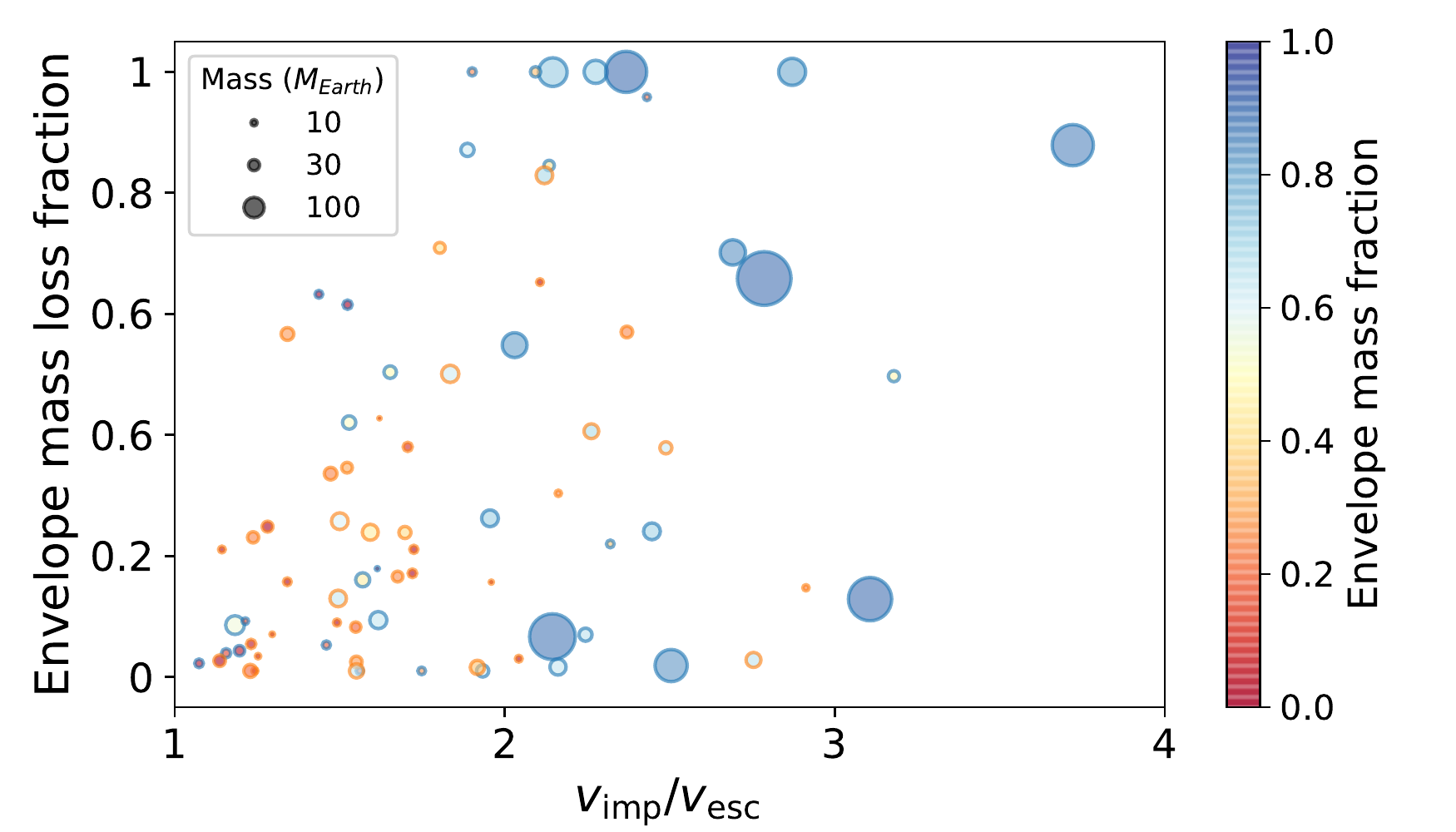}
\caption{Envelope mass-loss fraction for all the giant impacts in the simulations with the fiducial pebble flux. Circles with blue edges represent collisions in simulations without PEW effects, while those with orange edges are in simulations with PEW effects. Each size is proportional to mass, and the color corresponds to its envelope fraction before impacts.  A summary of all mass loss events is presented in Appendix~\ref{fig:app}.}
\label{fig:vimp_menv}
\end{figure}

Figure~\ref{fig:vimp_menv} shows the envelope mass loss fraction of each collision for all simulations of the fiducial flux in disks with and without PEW effects. It is shown that although the typical impact velocity is less than $2 \,v_{\rm esc}$, there exist some high-speed impacts with $> 2 \,v_{\rm esc}$. In some high-speed impacts, more than 90\% of the envelope can be lost. 
Notably, both the impact speed and the envelope loss fraction in the simulations in a disk without PEW effects are not lower than those in the simulations in a disk with PEW effects. Nevertheless, there is a large difference in the final envelope mass fraction between the simulations, as shown in Figure~\ref{fig:m_mcore}. This is because some accretions exist in the disk after the giant impacts ($t = 1-10 {\rm \,Myr}$) (bottom panel of Figure~\ref{fig:t_a}) in the disk with no PEW effects, and planets can reaccrete envelopes after giant impacts. Consequently, the final envelope mass fraction increases. Therefore, we can conclude that the rapid disk clearing after giant impacts plays a crucial role in the formation of giant planets with low envelope mass fractions.
We note that planets with low envelope mass fractions may form in the simulations with no PEW effects if further giant impacts occur after the disk gas is depleted ($t > 10 {\rm \,Myr}$). However, this is highly unlikely because the orbital separation between the planets is already large after the ``first'' giant impact phase ($t = 1-10 {\rm \,Myr}$). The system shown in Figure~\ref{fig:t_a}(a) is Hill stable \citep{1993Icar..106..247G}, and no further impact events would occur.

\begin{figure}[ht!]
\centering
\includegraphics[width=0.9 \columnwidth]{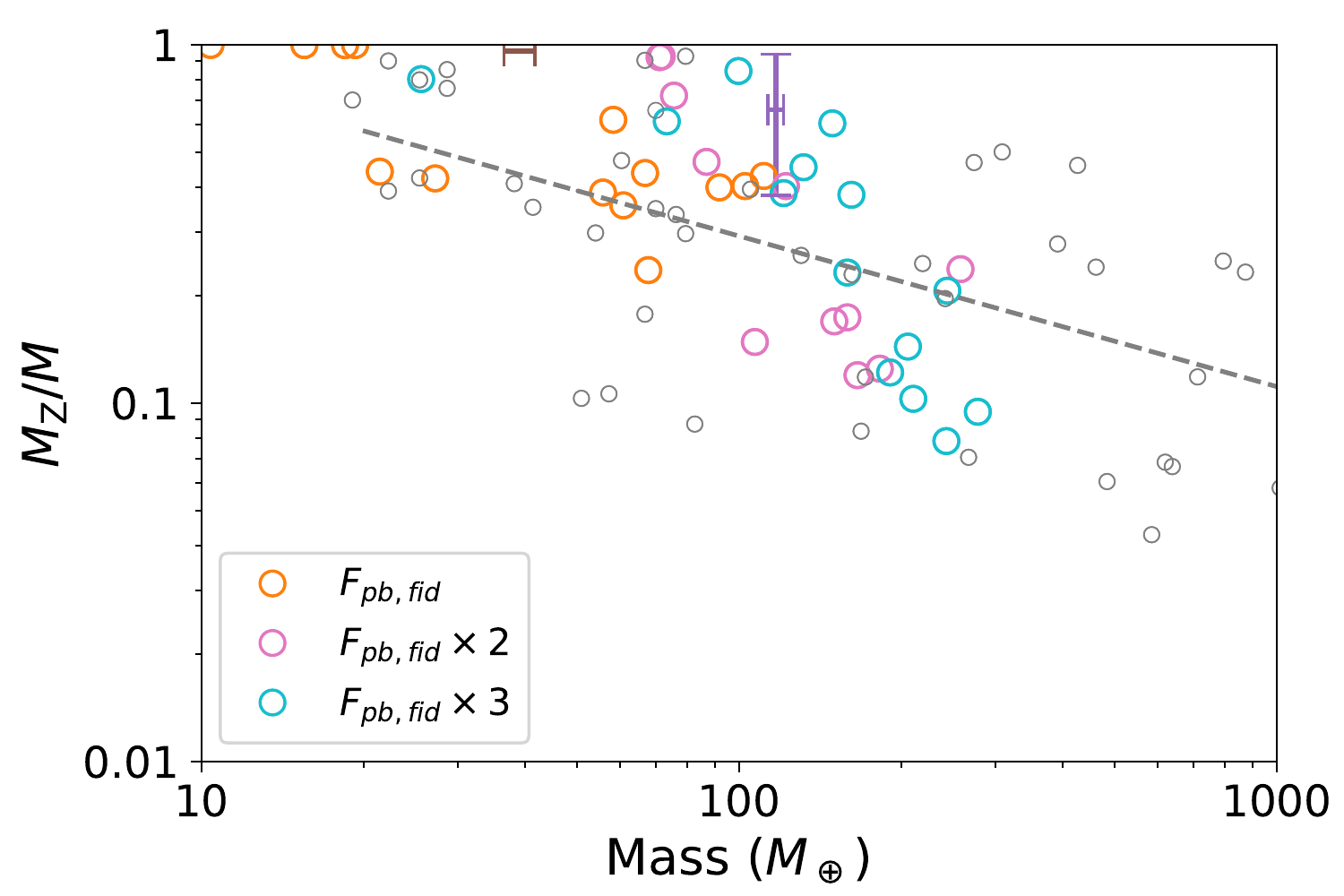}
\caption{Results of additional simulations in which the envelope loss during giant impacts is ignored.}
\label{fig:m_mcoref2}
\end{figure}

Next, we performed additional simulations to confirm the effect of impact erosion. Figure~\ref{fig:m_mcoref2} shows a summary of the simulations in which the mass loss during impacts was ignored. The other conditions were maintained as in the simulations shown in Figure~\ref{fig:m_mcore}(b). By comparing Figure~\ref{fig:m_mcoref2} with Figure~\ref{fig:m_mcore}(b), we find that the final metal-mass fraction is almost the same. In other words, even if the impact erosion is not considered, giant planets with high metal-mass fractions can form in a disk with PEWs. In contrast, by comparing the metal mass fraction in Figure~\ref{fig:m_mcoref2} with that in Figure~\ref{fig:m_mcore}(a), it is clear that the metal-mass fraction in Figure~\ref{fig:m_mcoref2} is significantly larger. Therefore, this result confirmed that the rapid disk clearing by PEWs was the main reason for the high metal-mass fraction.

Here, we comment on the model for the mass loss during impacts.
First, we do not consider the loss of metal materials during collisions. However, solid cores cannot merge completely when high-speed collisions occur \citep[e.g.,][]{2020MNRAS.496.1166D}. Even if such effects are considered, the planetary envelopes can be lost significantly, and the metal-mass fractions are expected to be large. Thus, our conclusions are likely to remain unchanged.
Second, it is not clear if the model used in this study is a good approximation because the model has not been confirmed for collisions between planets with high core masses ($>10\,M_\oplus$) and for hit-and-run collisions yet. Although these would not change our conclusions, detailed studies on the impact-induced envelope escape are still important.

\section{Conclusions}\label{sec:conc}

We demonstrate that the runaway gas accretion triggered by the formation of massive cores through giant impacts can be quenched by a rapid disk clearing due to PEWs, resulting in the formation of giant planets with high metal fractions. Further, we find that the metal mass in giant planets can increase after giant impacts between planets with the pebble isolation mass. 
The results reveal that giant planet collisions alone cannot explain the observed properties of transiting gas planets. 
The suppression of envelope accretion by the rapid-disk dissipation plays an important role in the increase in the metal-mass fraction of the giant planets.
As a result, the high metal mass and metal-mass fraction of the observed giant planets, including HD\,149026b and TOI-849b, can be naturally explained without relying on the impact erosion after the substantial depletion of the disk gas ($t > 10 {\rm \,Myr}$) and other envelope-loss mechanisms.

\begin{acknowledgements}
We thank the anonymous referee whose comments improved this manuscript.
Numerical computations were in part carried out on PC cluster at Center for Computational Astrophysics of the National Astronomical Observatory of Japan.
M.O. is supported by JSPS KAKENHI Grant Numbers 18K13608 and 19H05087; Y.H. by 18H05439; K.K. by 20J01258.
This study was inspired by discussions especially with Shigeru Ida and Daniel Thorngren at Ringberg Castle Workshop held in September 2019.
\end{acknowledgements}

\appendix
\section{A summary of all envelope mass loss events for Figure~\ref{fig:m_mcore}}
\label{sec:app1}

\begin{figure}[ht!]
\centering
\includegraphics[width=0.9\columnwidth]{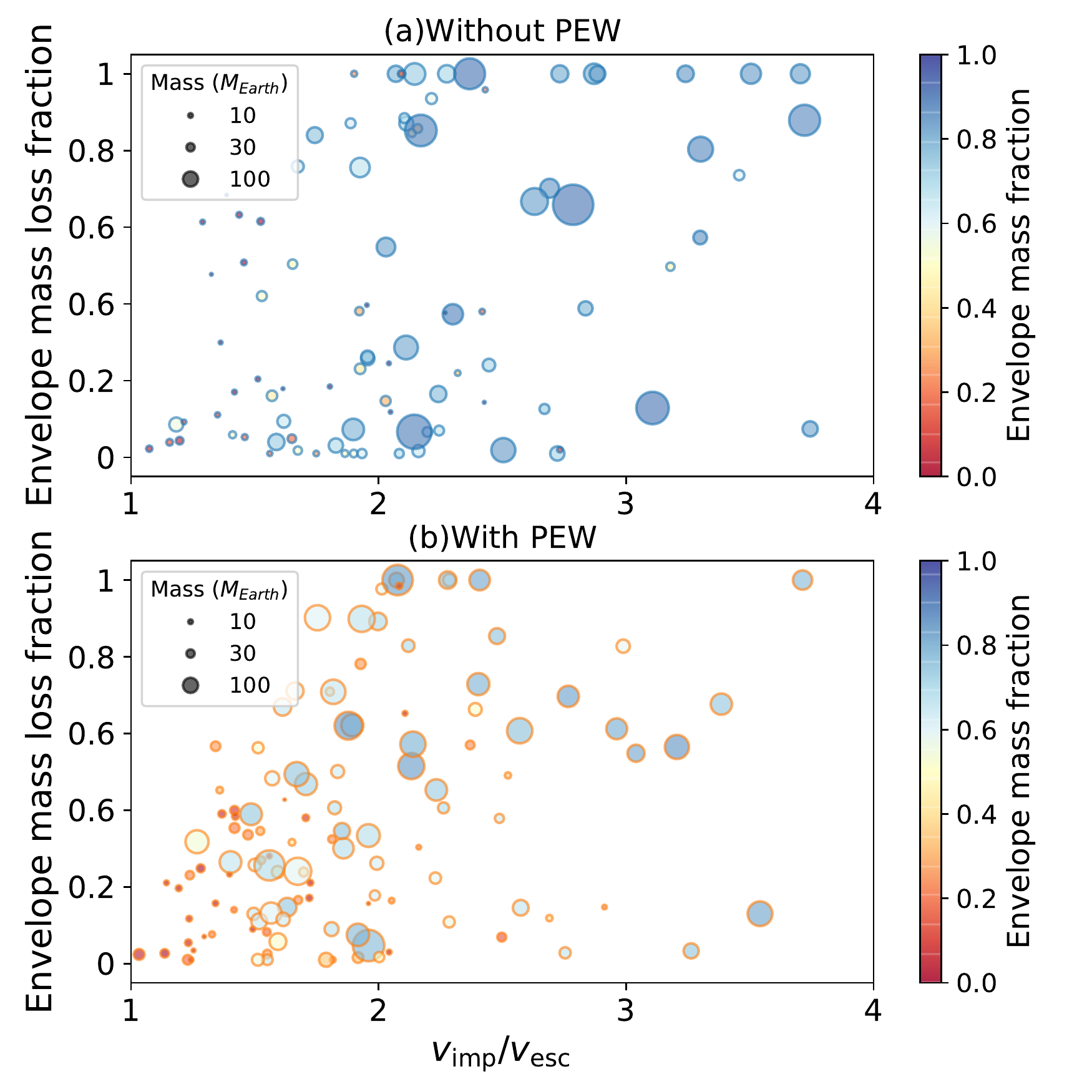}
\caption{Same as Figure~\ref{fig:vimp_menv}, but for all simulations for Figure~\ref{fig:m_mcore}.}
\label{fig:app}
\end{figure}

Figure~\ref{fig:vimp_menv} shows the envelope mass loss during giant impacts only for simulations with the fiducial pebble flux. Here, Figure~\ref{fig:app} summarizes the envelope mass loss for impact events in all simulations shown in Figure~\ref{fig:m_mcore}. As stated in Section~\ref{sec:erosion}, there are collisions in which a large fraction of the envelope is lost in both cases with and without PEW effects. Regarding the dependence on the planetary mass, we find that the impact velocity and/or the envelope loss fraction tend to increase when the mass is larger. Since the planetary mass and the envelope mass fraction is positively correlated (see Figure~\ref{fig:m_mcore}), the impact velocity tends to be higher for planets with high envelope mass fractions.

\bibliographystyle{aa}
\bibliography{reference}

\end{document}